\documentclass[12pt]{article}

\newcommand{\be}{\begin{equation}}
\newcommand{\ee}{\end{equation}}
\newcommand{\bea}{\begin{eqnarray}}
\newcommand{\eea}{\end{eqnarray}}

\newcommand{\bit}{\begin{itemize}}
\newcommand{\eit}{\end{itemize}}

\newcommand{\beqs}{\begin{eqnarray}}
\newcommand{\eeqs}{\end{eqnarray}}
\begin{document}
\bibliographystyle{h-physrev}
{\sf \title{Massive 3d Abelian gauge theories and electric-magnetic duality}
\author{Ansar Fayyazuddin\footnote{ansarf@aps.org}}
\maketitle
\begin{center}
\vspace{-1cm}
{\it American Physical Society \\ 1 Research Rd\\Ridge, NY 11961}
\end{center}

\begin{abstract}
We study a model of massive photons with a parity invariant  and non-local mass term.  We identify a discrete symmetry of the classical equations of motion and show that this symmetry can be thought of as an electric-magnetic duality valid only in the presence of a non-zero photon mass.  We explore this duality and discuss some features of the theory.  We define notions of electric and magnetic charge that can be computed in any spatial domain as a volume integral of a density.  These integrals can be transformed into integrals on the boundary.  For instance, the electric charge is given by the sum of the electric flux through the boundary of the domain and a line integral of the dual gauge field along that same boundary.  We show that electric and magnetic charge is quantized  $q_mq_e = 2\pi n m_\gamma/e^2$, where $e$ is the coupling constant and $m_\gamma$ is the mass of the photon.  Classical field configurations associated with external electric and magnetic charges as well as the classical forces between electric and magnetic charges are computed.
\end{abstract}

\vspace{-16cm}
\begin{flushright}
%HUTP-06/AXXX \\
%BCCUNY-HEP /06-XX \\
hep-th/yymmnnn
\end{flushright}

\thispagestyle{empty}

\newpage

\section{Introduction and summary}
Massive gauge bosons are of central importance to many lines of development in particle and condensed matter physics.  Yet it is not straightforward to give gauge bosons a mass.  The naive mass term, quadratic and local in the gauge field, $\int m^2A_\mu A^\mu$,  is not gauge invariant.  It gives a mass to the longitudinal component of the gauge field, which is neither gauge invariant nor has a kinetic term, and results in inconsistencies that cannot be cured. 

The best known mathematically consistent way to give gauge bosons a mass is the Higgs mechanism.  The Higgs mechanism requires the addition of new charged scalar degrees of freedom. If these charged scalar fields have a non-zero expectation value in the true vacuum of the theory, the gauge bosons that couple to the scalar fields acquire a mass and the would-be Goldstone modes are absorbed as the longitudinal components of the massive vector bosons.  In addition to the massive gauge bosons, there are typically massive scalar degrees of freedom left in the spectrum.  Therefore, the Higgs mechanism adds additional degrees of freedom to the spectrum beyond the desired massive vector bosons.     

In three dimensions, there is another way of giving gauge bosons a mass: through the introduction of a Chern-Simons term \cite{CS}.  The Chern-Simons term is a gauge-invariant but parity-violating term that results in a mass for the photon without the introduction of additional degrees of freedom.  We will have virtually nothing to say about this mechanism except for a passing remark in the concluding section of the paper.

The present paper is a study of aspects of a lesser known method of giving gauge bosons a mass \cite{Rob, AN, KKN-Mass, karsch, BP, JP}.  In this model of massive gauge bosons, one sacrifices locality but retains gauge and parity invariance.  Non-locality may strike one as an unacceptable feature but, as we shall see, the particular non-locality we need is of a mild variety and amounts to an ordinary mass term for the transverse gauge field.  The theory is perfectly local in the transverse, physical, polarizations of the photon.  The theory is also the effective theory derived in the context of hard thermal loops \cite{Rob, AN} and can be viewed as the effective field theory of the Abelian Higgs model, a manifestly local theory, in which the massive component of the Higgs field is arranged to have a much larger mass than the photon and is integrated out (or, more plainly, frozen to its classical expectation value).  

The main conclusion of this paper is that the classical theory of massive photons has an electric-magnetic duality that exchanges one massive photon for a different one.  We explore the consequences of this duality and show that two equivalent descriptions of the theory exist in which the electric sources of one theory are magnetic sources of the other.  We then show that in any given domain, the electric charge can be computed by an integral at the boundary of the domain.  The integral is the sum of the flux of the electric field through the boundary and a line integral of the dual gauge field along the boundary.  A similar result holds for the magnetic charge.  Interestingly, if the domain is taken to be the infinite plane, the charge of localized sources is obtained entirely from the line integral of the dual gauge field (the electric-field flux vanishes at infinity as one would expect for a massive photon) at infinity.  We also show that charge is quantized by considering a quantum magnetic test particle in the presence of an electric source.  

\section{Massive Abelian gauge theory}

This paper is concerned with the following classical theory:
\be
S = -\int d^3x\frac{1}{4e^2}F_{\mu\nu}F^{\mu\nu} - \frac{m^2}{e^2}\Gamma[A] \label{action}
\ee
where $\Gamma$ satisfies
\be
\Gamma(A_\mu+\delta A_\mu) = \Gamma (A_\mu) + \int d^3x l^{\mu}(x)\delta A_\mu(x) + \hbox{O($\delta A_\mu^2$)}
\ee
and $l_{\mu}(x)$ is defined through the relations:
\bea
&&\partial_\mu l_\nu - \partial_\nu l_\mu = F_{\mu\nu} \label{def}\\
&&\partial_\mu l^\mu = 0 \label{inv}.
\eea
The first of the above equations, (\ref{def}), implies 
\be
l_{\mu}(x) = A_\mu + \partial_\mu \lambda
\ee
The second equation, (\ref{inv}), determines $\lambda$ to yield 
\be 
l_\mu  = A_\mu -  \partial_\mu\frac{1}{\partial_\alpha \partial^\alpha}\partial^\nu A_\nu.
\ee  
Thus $l_\mu$ is the gauge invariant transverse gauge field or, equivalently, the gauge field $A$ with its longitudinal component projected out.  The mass term, $\Gamma [A]$, can be written down explicitly and appears below in equation (\ref{Gamma}).

The equations of motion of this theory expressed in terms of the gauge field strength and $l$ are:
\be
\partial^\mu F_{\mu\nu} - m^2l_\nu = 0
\ee
Using (\ref{def}) this can be rewritten as:
\bea
0 &=& \partial^\mu ( \partial_\mu l_\nu - \partial_\nu l_\mu) - m^2l_\nu \nonumber \\
&=&  (\partial^\mu \partial_\mu - m^2) l_\nu
\eea
where the second equality follows from (\ref{inv}).   Thus $l_\mu$, the transverse gauge field, satisfies the Klein-Gordon equation with mass $m$. 

Next, we define the Hodge dual of $F$:
\be
L_\mu = \frac{1}{2}\epsilon_{\mu\nu\rho}F^{\nu\rho},
\ee
which can be used to reexpress the Bianchi identity and equation of motion, respectively, as follows:
\bea
\partial_\mu L^\mu &=& 0 \\
\partial_\mu L_\nu - \partial_\nu L_\mu &=& -m^2 \epsilon_{\mu\nu\rho}l^\rho.
\eea
The complete set of equations of motion and identities are:
\bea
\partial_\mu L^\mu &=& 0 \\
\partial_\mu l_\nu - \partial_\nu l_\mu &=& F_{\mu\nu} \\
\partial_\mu l^\mu &=& 0 \\
\partial_\mu L_\nu - \partial_\nu L_\mu &=& -m^2 \epsilon_{\mu\nu\rho}l^\rho 
\eea
These equations can be expressed compactly as:
\bea
d*L &=& 0 \nonumber \\ 
dl + *L&=&0 \nonumber \\
d*l &=& 0 \label{eom}\\
dL + m^2 *l&=& 0 \nonumber
\eea
Note that the first three equations are identities and only the last equation is a true equation of motion.  

\section{A duality}
The above equations are invariant under the following transformation:
\bea
l\rightarrow \frac{1}{m}L \nonumber \\ \label{duality}
L\rightarrow m l
\eea
This symmetry of the equations of motion presumes that the mass is nonzero.  Note that the odd appearance of the mass is consistent with the fact that $L$ and $l$ have different mass dimensions.

The transformation (\ref{duality}) is not an invariance of the action, but rather the action picks up a minus sign under it.   To see this more explicitly, we first write down $\Gamma(A)$ in closed form.  As noted above
\be
l_\mu = A_\mu - \partial_\mu\frac{1}{\partial^2}\partial_\nu A^\nu \equiv P_\mu^\nu A_\nu
\ee 
where $P_\mu^\nu = \delta_\mu^\nu - \frac{1}{\partial^2}\partial_\mu\partial^\nu$ is a projection operator on to the transverse part of the gauge field.  It satisfies the usual defining propoerty of a projection operator: $P_\mu^\nu P_\nu^\rho = P_\mu^\rho$.  It follows then that $\delta l_\mu = P_\mu^\nu \delta A_\nu$ and:
\bea
&&\delta (l_\mu l^\mu) \nonumber \\
&=& 2\delta A^\mu l_\mu + O(\delta A)^2
\eea
Thus $\Gamma(A)$ from (\ref{action}) can be written explicitly as:
\be
\Gamma (A) =\int d^3 \frac{1}{2}l_\mu l^\mu \label{Gamma}
\ee
and the complete action is:
\bea
S = \int d^3x[\frac{1}{2e^2}L_{\mu}L^{\nu} - \frac{m^2}{2e^2}l_\mu l^\mu] \label{compaction}
%&=& \int d^3x[-\frac{1}{4e^2}F_{\mu\nu}F^{\mu\nu} - \frac{m^2}{2e^2}(A_\mu - \partial_\mu\frac{1}{\partial^2}\partial_\nu A^\nu)(A^\mu - \partial^\mu\frac{1}{\partial^2}\partial_\rho A^\rho)] \\ 
\eea
It is clear that under (\ref{duality}), the action is only invariant up to an overall minus sign.  This feature - the non-invariance of the action - is shared by the electric-magnetic duality of Maxwell theory in 4 dimensions.  A discussion of this point in the 4D theory and the construction of a duality invariant action can be found in \cite{ss}.

Since the transformation (\ref{duality}) is only valid when $m\neq 0$, it is convenient to redefine the fields so that $l$ and $L$ have the same dimension.  Rescaling $l\rightarrow ml$, the equations  (\ref{eom}) become:
\bea
d*L &=& 0 \nonumber \\ 
dl + m*L&=&0 \nonumber \\
d*l &=& 0 \label{neweom}\\
dL + m *l&=& 0 \nonumber
\eea
and the transformation (\ref{duality}) takes the simpler form:
\bea
l\rightarrow L \nonumber \\ \label{newduality}
L\rightarrow l.
\eea

In the remainder of this section, we will explore the meaning of this symmetry of the equations of motion and argue that it is, as far as we are aware, a new electric-magnetic duality valid only for massive gauge theories.  

In the massless case, electric-magnetic duality refers to the exchange of the Bianchi identities with the equations of motion.  This can be expressed as exchanging the field strength $F$ for its Hodge dual $*F$.  In $2+1$ dimensions, under this duality, the massless gauge field is exchanged for a massless scalar field $\phi$ with $F=*d\phi$.  With this identification, what were previously called the equations of motion are identically satisfied, while what was the Bianchi identity has to be imposted as an equation of motion for the scalar field $\partial^\mu\partial_\mu \phi = 0$.  This is all well known.

We would like to proceed by analogy with the massless case.  We thus separate the equations (\ref{neweom}) into identities and equations of motions.  As already mentioned, there are three identities 
\bea
d*L &=& 0 \nonumber \\ 
dl + m*L&=&0 \label{Aidentities}\\
d*l &=& 0.  \nonumber
\eea
The three equations that constitute (\ref{Aidentities}) are respectively the Bianchi identity and the defining equations for $l$.  
The equation of motion is
\be
dL + m *l= 0. \label{Aeom}
\ee
We will call this description "Theory A". 
The identities (\ref{Aidentities}) can be solved in terms of a vector field $A$:
\bea
*L &\equiv& -F = -dA \label{A}\nonumber \\
l_\mu &=& m(A_\mu - \frac{1}{\partial^2}\partial_\mu\partial_\nu A^\nu) 
\eea
and the equation of motion can be derived from the previously discussed action:
\be
S_A = \int d^3x[\frac{1}{2e^2}L_{\mu}L^{\nu} - \frac{1}{2e^2}l_\mu l^\mu]
\ee
where the label $A$ has been added to the action to indicate that the theory is to be understood in terms of the gauge field as given in (\ref{A}).

We will now define the dual theory, "Theory B", as the identities:
\bea
d*l &=& 0 \nonumber \\ 
dL + m*l&=&0 \label{Bidentities}\\
d*L &=& 0  \nonumber
\eea
and the equation of motion:
\be
dl + m *L= 0 \label{Beom}
\ee
Just as with Theory A, the first of the identities in (\ref{Bidentities}) serves as a Bianchi identity and the remaining two equations define $L$.  As with Theory A, the identities (\ref{Bidentities}) can be solved in terms of a new vector field $a$:
\bea
*l &\equiv& -f = -da \label{a}\nonumber \\
L_\mu &=& m(a_\mu - \frac{1}{\partial^2}\partial_\mu\partial_\nu a^\nu) 
\eea
and the equation of motion can be derived from the action:
\be
S_B= \int d^3x[\frac{1}{2e^2}l_{\mu}l^{\nu} - \frac{1}{2e^2}L_\mu L^\mu]
\ee
The label $B$ indicates that the theory is to be understood in terms of the gauge field $a$ as given in (\ref{a}).  

\section{Static sources, electric and magnetic charges, and test particles}
The terms "electric" and "magnetic" have specific meanings that are independent of what we have chosen to call electric-magnetic duality.  Namely, electric charges produce  electric fields ($E_i = F_{0i} = -\epsilon_{ij}L_j$), while magnetic charges produce a magnetic field $B = F_{12} = L_0$.  We turn now to static sources that couple to the gauge fields of Theory A and Theory B.  

\subsection{Static sources}

Consider a static external source $J= q_e\delta^{(2)}({\bf r})dt$ in Theory A, $\delta S_A =\int d^3xA_\mu J^\mu$.   The coupling to the gauge field $A_\mu$ is such that $J$ represents an electric source of Theory A.   With the introduction of the source, the equation of motion (\ref{Aeom}) changes to:
\be
dL + m *l= e^2*J. \label{Aeoms}
\ee
or expressed in terms of the gauge field and it's associated field strength:
\bea
-\partial^iF_{0i} &+&m^2(A_0- \partial_0\frac{1}{\partial^2}\partial_\nu A^\nu) = -e^2q_e\delta^{(2)}(r) \nonumber \\
\partial_i F_{12} &+&m^2(A_i-\partial_i \frac{1}{\partial^2}\partial_\nu A^\nu) = 0
\eea
For a static source, we can take $A_\mu$ to be independent of time and the equations can be solved, in analogy with Maxwell theory, by setting $A_0=\phi, A_i=0$:
\be
\nabla^2\phi - m^2\phi = e^2q_e\delta^{(2)}(r).
\ee
The above equation is solved by:
\be
\phi (r)= \frac{e^2q_e}{2\pi} K_{0}(mr),
\ee
where $K_0$ is the zeroth modified Bessel function of the second kind.  Note that the only non-zero component of the field strength is the electric field:
\bea
F_{0i} &=& -\partial_i\phi = \frac{e^2q_e}{2\pi}mK_1(mr) \frac{x^i}{r}\nonumber \\
F_{12} &=& 0 \label{statelec}
\eea
as one would expect for a static electric source.  In the above equation, we used the relation $K_0' = -K_1$, where $K_1$ is the modified Bessel function of the second kind of order 1.  

We now express the above solution in terms of the fields of Theory B.  Since $l_0$ is the only non-zero component of $l$, the field strength associated with the gauge field $a$ is 
\bea
f_{12}&=& l_0 = \phi = \frac{e^2q_e}{2\pi}K_{0}(mr) \nonumber \\
f_{0i}&=& 0. 
\eea
Thus we find that the static {\em electric} source in Theory A produces a {\em magnetic} field in terms of the fields ofTheory B and {\em no} electric field.  We conclude, therefore, that electric charges in Theory A are magnetic charges in Theory B and vice versa.

Of course, an identical set of steps can be carried out for a static source that couples minimally to the gauge field, $a_\mu$, in Theory B: the source $j= q_m\delta^{(2)}(r)dt$ couples to $a_\mu$ through the term in the action $\delta S_B =\int d^3xa_\mu j^\mu$.  The equation of motion is solved by:
\bea
a_i &=& 0 \nonumber \\
a_0 \equiv \psi(r) &=&  \frac{e^2q_m}{2\pi} K_{0}(mr)
\eea
The field strengths associated with this configuration in Theory A and Theory B can be summarized as follows:
\bea
f_{0i} &=& -\partial_ia_0 \nonumber \\
f_{12} &=& 0 \label{statmag}
\eea
as one would expect for a static electric source.
Expressed in terms of the fields of Theory A, the field strength associated with the gauge field $A$ the solution is 
\bea
F_{12}&=& L_0 = \psi \nonumber \\
F_{0i}&=& 0.
\eea

Thus, we have established that electrically charged particles of Theory A are magnetically charged in Theory B and {\em vice versa}.

\subsection{Computing Charges}
In Maxwell theory, the electric charge contained in any domain $D$ is computed by the integral:
\be
Q[D] = \int_D d^2x {\bf \nabla}\cdot{\bf E} = \int_{\partial D} {\bf E}\cdot d{\bf S} 
\ee
where in the second step we used the well-known conversion of the volume integral of the divergence of the electric field to its flux through the boundary of $D$, $\partial D$.  From the considerations in the previous subsection, we can define a similar quantity in Theory A:
\be
Q[D] = \int_D d^2x {\bf \nabla}\cdot{\bf E} - m\int_D l_0 d^2x  \label{Acharge}
\ee
The first integral can again be converted, as before, to a surface integral at the boundary of $D$.  What about the second term?  It too can be converted to a surface integral by noting that $l_0$ can be expressed as
\be
l_0 = f_{12} = \partial_1 a_2 - \partial_2 a_1
\ee
Thus 
\be
\int_D l_0 d^2x = \int_D d_2a = \oint_{\partial D}a.
\ee
The charge can then be computed as 
\be
Q[D] = \int_{\partial D} {\bf E}\cdot d{\bf S} - m\oint_{\partial D}a \label{Acharge2}
\ee

In analogy with the above considerations, a similar quantity can be defined in Theory B:
\bea
q[D] &=&  \int_D d^2x {\bf \nabla}\cdot{\bf e} - m\int_D L_0 d^2x \nonumber \\
&=& \int_{\partial D} {\bf e}\cdot d{\bf S} - m\oint_{\partial D}A \label{Bcharge}
\eea
where $e_i = f_{0i}$.   

As an example we can compute the charge of the static solutions of the previous subsection.  In the case of the static electric charge in Theory A, the electric field, $E_i = -\partial_i\phi$.  In order to calculate the charge, we need to compute two integrals in either form (\ref{Acharge}) or (\ref{Acharge2}).  If we want to use the second form (\ref{Acharge2}), we need to find $a_\mu$ first.  It is straightforward to check that 
\be
a = \frac{e^2q_e}{2\pi m}(mrK_1(mr)-1)(\frac{x}{r^2}dy - \frac{y}{r^2}dx) = \frac{e^2q_e}{2\pi m}(mrK_1(mr)-1)d\theta \label{a-statelec}
\ee 
For simplicity we will compute $Q[D]$ for $D=\{(x,y): x^2+y^2 < r^2\}$ and $\partial D\{(x,y): x^2+y^2 =r^2\}$:
\bea
Q[D] &=& E_r (r)r2\pi - m a_\theta(r)2\pi \nonumber \\
&=& \frac{e^2q_e}{2\pi}mK_1(mr)2\pi r - m\frac{e^2q_e}{2\pi m}(mrK_1(mr)-1) 2\pi \\
&=& e^2q_e
\eea
This charge is independent of $r$. It is interesting to note that as $r\rightarrow\infty$, only the second term contributes to the charge since $K_1$ vanishes exponentially.  
We can also compute the magnetic charge $q[D]$ for the same field configuration.  Since $a = a_\theta(r)d\theta$, $e=0$ and $A = A_0dt$, 
\be
q[D] = 0.
\ee

As noted in the previous paragraph, when the electric charge is computed on the infinite plane, the electric-field flux contribution to the charge vanishes at infinity but the line integral of the dual gauge field does not and alone accounts for the entire charge.  The surprise here is that while one expects the charge in the bulk to be screened due to the photon mass, it nevertheless can be "seen" at infinity via the line integral of the dual gauge field.  One may wonder if there are implications for reconciling confinement with screening in non-Abelian theories with effectively massive gluons\footnote{I owe this observation entirely to V. P. Nair.}.

\subsection{Test charges}
Next we consider introducing test charges into an arbitrary field configurations.  There are two obvious types of test particles: ones that couple minimally to $A_\mu$, or "electric" particles:
\be
S_e = m\int ds + Q\int A_\mu \frac{dx^\mu}{ds}ds
\ee
and those that couple minimally to $a_\mu$, or "magnetic" particles:
\be
S_m = m\int ds + q\int a_\mu \frac{dx^\mu}{ds}ds.
\ee
In the electric test particle case, the equation of motion is:
\be
\frac{dp^i}{dt} = Q(E_i + \epsilon_{ij}v_jB)
\ee
while for the magnetic test particle case:
\be
\frac{dp^i}{dt} = q(e_i + \epsilon_{ij}v_jb).
\ee

By using the expressions for the electric field produced by a static electric source given in (\ref{statelec}), we can find the "Coulomb" force law by using the above equations of motion for a test electric particle:
\be
\frac{dp^i}{dt} = Q\frac{e^2q_e}{2\pi}mK_1(mr) {\bf \hat r}
\ee
Two magnetic charges also have the same force law with magnetic charges replacing the electric ones in the above expression.  We can also find the equation of motion of a magnetic charge in the presence of a static electric source:
\be
\frac{dp^i}{dt} = q\epsilon_{ij}v_jb = q\frac{e^2q_e}{2\pi}K_{0}(mr) \epsilon_{ij}v_j
\ee
where $v_i$ is the velocity vector of the magnetic test charge.

\subsection{Charge quantization}
So far we have discussed a version of massive electrodynamics in purely classical terms.  Without attempting to quantize the massive photon theory, we can consider the quantum mechanics of a test particle in the gauge theory background.  In particular, let us consider a magnetically charged particle in the background of a charged static electric source.  That is, we are considering the quantum mechanics of a test particle coupled minimally to $a_\mu$, where  $a_\mu$ is given by the expression (\ref{a-statelec}).  

When the test magnetic charge is transported along a closed path $C$ around $r=0$, the location of the static electric charge, the wavefunction of the test magnetic charge, $q_m$, picks up a phase:
\bea
\exp iq_m\oint_C a &=& \exp iq_m\frac{e^2q_e}{2\pi m}(mrK_1(mr)-1)d\theta \nonumber \\
&=& \exp iq_m\frac{e^2q_e}{2\pi m}(mrK_1(mr)-1)2\pi 
\eea
If we take $C$ to be at infinity, the phase:
\be
\exp iq_m\oint_{C_\infty} a = \exp -iq_m\frac{e^2q_e}{m} 
\ee
If we require that at infinity the phase factor is $1$, we have a quantization condition on the product of electric and magnetic charges allowed in the theory:
\be
q_mq_e = \frac{m}{e^2}2\pi n 
\ee
where $n$ is an integer.

\section{Conclusions}
In this paper we made some observations about a particular classical theory of massive electrodynamics in $2+1$ dimensions.  We showed that the classical equations of motion enjoy a discrete symmetry that amounts to an electric-magnetic duality which differs from the usual notion of electric-magnetic duality for the massless theory.  In the usual massless case, electric-magnetic duality exchanges the massless Abelian gauge field for a massless scalar field.  In the case studied here, the dual theory is also that of a massive Abelian gauge field.    

Another difference between the duality presented here and conventional electric-magnetic duality valid for massless photons is that in the massless case the dual descriptions are not simultaneously valid.  Thus, in $2+1$ dimensions, when sources are present, either the scalar field description or the gauge field description is valid, not both (at least globally).  This is because the Bianchi identity of at least one description fails in regions where a source is non-vanishing.  This is not the case here.  The fact that we can use either description relies on the equations:
\bea
\partial_\mu l^\mu &=& 0 \nonumber \\
\partial_\mu L^\mu &=& 0, \label{bi}
\eea
which are valid even in the presence of sources.   

One could imagine that the natural theory of a massive photon would be a mass term deformation of the dual massless scalar theory.  A mass term for the scalar theory is quite natural and does not pose any obvious conceptual problems.  However, the  relationship of a massive scalar to a vector field theory is not clear.  The considerations presented above would seem to suggest that the massless and massive Abelian gauge theories are not smoothly connected.  

An intriguing feature of the theory studied here is that of the quantization of charge.  In the massless case, where there is only a single dimensionful parameter (the gauge coupling), there is no obvious way for the quantization of charge to occur.  However, in the presence of the photon mass, a dimensionless ratio can be constructed and it is interesting that this ratio provides the basis for charge quantization.  

Finally, we mention a few questions that may be interesting to pursue in the future.  In addition to the non-local mass term that we studied, one could consider adding a Chern-Simons term.  The Chern-Simons action takes a particularly natural form in terms of the fields that we have introduced\footnote{I thank V. P. Nair and Abhishek Agarwal for this insight.}:
\be
S_{CS} = k\int A\wedge dA = \frac{k}{m}\int l_\mu L^\mu.
\ee
The consequences for the duality considered here in the presence of the Chern-Simons term would be an interesting question to study.  Although we will not pursue this question in depth here, we briefly note that the equations for the theory consist of the identities (\ref{bi}) and:
\bea
dl -ke^2 l + m*L =0 \nonumber \\
dL -ke^2 L + m*l  = 0.
\eea
These equations are invariant under the electric-magnetic duality transformation $l \leftrightarrow L$.
A second natural question to consider is how the narrative presented here would be modified in the non-Abelian Yang-Mills case.  We leave these questions for the future.

{\bf Acknowledgements:} \\
I am grateful to Abhishek Agarwal and V. Parameswaran Nair for invaluable discussions, for their generosity in sharing their many insights and for encouraging me to write up these notes.  I thank Rob Pisarski for discussions and his comments on the paper.

\end{document}